# Ultrafast two-dimensional field spectroscopy of terahertz intersubband saturable absorbers

**JÜRGEN RAAB,[1] CHRISTOPH LANGE,[1] JESSICA L. BOLAND,[1] IGNAZ LAEPPLE,[1] MARTIN FURTHMEIER,[1] ENRICO DARDANIS,[2] NILS DESSMANN,[2] LIANHE LI,[3] EDMUND H. LINFIELD,[3] A. GILES DAVIES,[3] MIRIAM S. VITIELLO,[2,4] AND RUPERT HUBER[1,5]**

[1]*Department of Physics, University of Regensburg, 93040 Regensburg, Germany*
[2]*NEST, CNR-Istituto Nanoscienze and Scuola Normale Superiore, Piazza San Silvestro 12, Pisa I-56127, Italy*
[3]*School of Electronic and Electrical Engineering, University of Leeds, Leeds LS2 9JT, UK*
[4]*miriam.vitiello@sns.it*
[5]*rupert.huber@ur.de*

**Abstract:** Intersubband (ISB) transitions in semiconductor multi-quantum well (MQW) structures are promising candidates for the development of saturable absorbers at terahertz (THz) frequencies. Here, we exploit amplitude and phase-resolved two-dimensional (2D) THz spectroscopy on the sub-cycle time scale, to observe directly the saturation dynamics and coherent control of ISB transitions in a metal-insulator MQW structure. Clear signatures of incoherent pump-probe and coherent four-wave mixing signals are recorded as a function of the peak electric field of the single-cycle THz pulses. All nonlinear signals reach a pronounced maximum for a THz electric field amplitude of 11 kV/cm and decrease for higher fields. We demonstrate that this behavior is a fingerprint of THz-driven carrier-wave Rabi flopping. A numerical solution of the Maxwell-Bloch equations reproduces our experimental findings quantitatively and traces the trajectory of the Bloch vector. This microscopic model allows us to design tailored MQW structures with optimized dynamical properties for saturable absorbers that could be used in future compact semiconductor-based single-cycle THz sources.

## 1. Introduction

Terahertz photonics has emerged as an exciting field of research owing to its enormous promise in a fast-growing variety of applications. These range from ultrafast spectroscopy and sub-cycle control of condensed matter to non-invasive imaging, chemical fingerprinting and security screening [1,2]. To address these applications, rapid progress has been made in the development of advanced table-top THz sources, employing a variety of generation schemes. For example, photoconductive antennas utilize resonantly excited, DC-biased semiconductor materials to generate single-cycle, carrier-envelope phase stable THz pulses [3]. Optical rectification of ultrashort near-infrared (NIR) laser pulses in nonlinear $X^{(2)}$ crystals [4] or air plasmas [5] covers the entire THz to NIR window. It also allows THz carrier waves to reach atomic field strengths, which can be used to drive extreme nonlinearities, such as high-harmonic generation [6], coherent spin control [7], and multi-wave mixing [8,9]. Free-electron lasers can also be widely tuned over the whole THz spectrum, whilst maintaining high output power [10]. Yet, the above sources typically depend on expensive femtosecond laser sources, limiting their use in cost-effective applications outside specialized optics laboratories.

Electrically-pumped quantum cascade lasers (QCLs) offer a promising alternative. They exploit intersubband transitions in semiconductor heterostructures, enabling a wide operational frequency range through bandstructure engineering. QCLs can also operate as optical frequency combs and provide high output powers [11-17]. Despite these advantages, ultrashort pulse generation in broadband QCLs can be only achieved through active mode-locking, modulation of the QCL gain by an external radio frequency signal, or through phase synchronization [18-21]. In order to further miniaturize QCL-based THz sources towards on-chip products, passively mode-locked QCLs have been suggested. However their realization requires custom-tailored THz saturable absorbers, whose coherent and incoherent characteristics on femtosecond timescales need to be precisely determined.

In the visible to mid-infrared spectral region, semiconductor-based saturable absorbers have been routinely used to achieve passive mode-locking [22]. In these schemes, the high peak intensity of pulsed light suffers substantially lower losses within the saturable absorber than the cw background, which allows for a robust formation of ultrashort pulses in the cavity. Saturable absorption has also been shown in the THz frequency range using n-doped bulk semiconductors [23]. However, problems still arise when such systems are integrated into a laser cavity, due to the inevitable losses by free-carrier absorption and large saturation intensities. An alternative approach to the development of THz saturable absorbers is the use of single and few layers of graphene [24], which have shown very low saturation intensities, on the order of a few W/cm²; there is, however, a lack of flexibility in the design parameters such as the saturation threshold and relaxation time. Saturable absorbers exploiting MQW semiconductor heterostructures provide a valuable alternative. Specifically, intersubband (ISB) transitions in MQWs allow for the customized design of absorption frequency, recovery times, and saturation intensities. In addition, semiconductor quantum wells can be readily integrated into laser cavities, facilitating their uptake. While the concept of ISB transitions is well known and the dynamics of their interaction with light have been studied in great detail [25-30], their coherent and incoherent nonlinear dynamics in the THz frequency range have yet to be evaluated.

## 2. Field-sensitive 2D THz spectroscopy of ISB transitions

Here, we use field-sensitive 2D spectroscopy to investigate the ultrafast nonlinear response, as well as the saturation dynamics, of an ISB transition in a MQW system, which is coupled to a THz electromagnetic field via a metallic grating. Phase-stable THz pulses coherently excite and probe both the ISB population and polarization with sub-cycle temporal resolution. Electro-optic sampling of the transmitted waveform allows for amplitude- and phase-sensitive observation of incoherent pump-probe and coherent four-wave mixing signals. The intensity dependence of these nonlinearities exhibits the hallmark of THz carrier-wave Rabi flopping, quantitatively confirmed by a numerical solution of the Maxwell-Bloch equations. The resulting dynamic modulation and saturation of the absorption on the time scale of a single optical cycle makes them an ideal component for developing few-cycle or even single-cycle THz QCLs.

The MQW structure (sample code G0061) investigated in this work consists of 35 GaAs quantum wells, each of thickness 36 nm, separated by 20-nm-thick $Al_{0.15}Ga_{0.85}As$ barriers, which are delta doped at a density of $5 \times 10^{10}$ cm$^{-2}$ after the first 5 nm of the barrier is grown (see Fig. 1(a)). The resulting bandstructure, calculated by a Schrödinger-Poisson solver [31], is shown in Fig. 1(b), together with the numerically calculated wave functions of the first and second subbands. The transition energy between the two subbands corresponds to a frequency of $\nu_{ISB}$ = 2.7 THz (see transmission spectrum in Fig. 1(e)) and the transition dipole moment, calculated from the simulated wave functions, is $\mu_{12}/e$ = 3.8 nm, where $e$ is the elementary charge. This value for the dipole moment already suggests low saturation thresholds (of ~ several kV/cm), highlighting the potential of the MQW structure for realizing passively mode-locked THz QCLs. Since, however, the dipole is oriented out of the quantum well plane, THz radiation incident in the growth direction cannot couple to the ISB transition. In order to resolve this problem and to decrease the necessary saturation threshold further, a gold grating (lattice period: 16 µm) was structured on top of the MQW system by means of a combination of UV lithography, metal evaporation and lift-off. The grating design was guided by finite-difference Fourier-domain simulations to provide both strong field confinement and the presence of a z-polarized component in the near-field region (see Fig. 1(c)), which then couples directly to the dipole moment of the ISB transition.

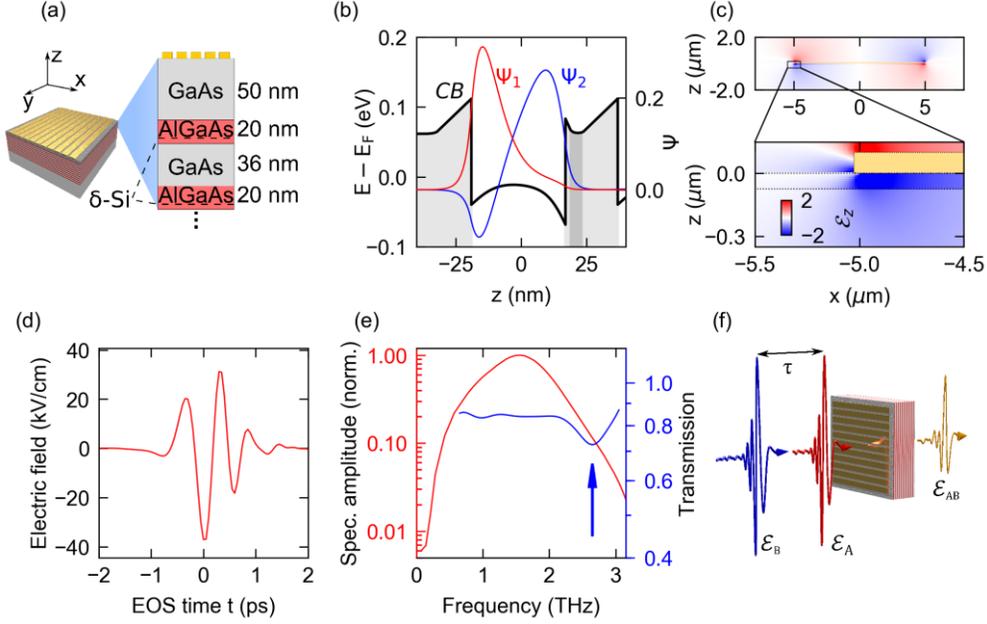

Fig. 1. (a) Schematic diagram of the THz saturable absorber structure showing the grating and the MQW stack. δ-Si: Silicon delta-doping layers. (b) Electron envelope functions of the first ($\Psi_1$, red) and second ($\Psi_2$, blue) subbands, and the conduction band edge (CB, black), in the MQW structure. (c) Cross section of the sample, showing the simulated field enhancement of the z-component $\mathcal{E}_z$ at 2.7 THz underneath one period of the gold grating, normalized to the incident electric field. Dashed horizontal lines indicate a GaAs layer, separating the MQW section from the metal grating. Lower panel: magnified view of the marked part of the upper panel. (d) Electric field waveform of the THz pulses used to excite the ISB system. (e) Amplitude spectrum of the THz transient shown in (d) along with the measured field transmission of the sample. The blue arrow indicates the expected ISB transition frequency. (f) Experimental principle showing the two identical THz pulses with fields $\mathcal{E}_A$ and $\mathcal{E}_B$ delayed by a time $\tau$, which prepare and interrogate the structure's nonlinear response.

To investigate the dynamics of this saturable absorber in the most direct and comprehensive way, we perform 2D phase- and amplitude-resolved THz high-field spectroscopy [32-34]. To optimize the signal-to-noise ratio, we developed a dedicated high-repetition-rate spectroscopy system based on an Yb:KGW amplifier that generates 260-fs pulses centered at a wavelength of 1028 nm (repetition rate: 50 kHz). Intense, phase-stable THz transients (see Fig. 1(d) and 1(e)) are generated in a LiNbO$_3$ crystal via optical rectification of the near-infrared laser pulses in a tilted pulse-front scheme [4]. In a Michelson interferometer, these THz transients are split into two identical pulses, A and B, which can be delayed with respect to each other with a variable delay time $\tau$ (see Fig. 1(f)). The two pulses are then focused onto the cryogenically-cooled sample to study the nonlinear response of the saturable absorber system. The total THz electric field transmitted through the sample is finally detected in amplitude and phase by electro-optic sampling in a 500 µm ZnTe crystal, as a function of both the electro-optic sampling time $t$, and the delay time $\tau$. By chopping the THz radiation in both arms of the Michelson interferometer at different subharmonic frequencies of the laser repetition rate, the transmitted field for all possible chopper combinations is obtained. Thereby, we extract the field $\mathcal{E}_{AB}$, from both pulses and the contributions $\mathcal{E}_A$ and $\mathcal{E}_B$ from each individual pulse. Consequently, the correlated nonlinear response $\mathcal{E}_{NL}$ can be determined by subtracting the contributions from each of the single pulses from the total field, i.e., $\mathcal{E}_{NL} = \mathcal{E}_{AB} - \mathcal{E}_A - \mathcal{E}_B$. Figure 2(a) depicts $\mathcal{E}_{NL}$ for a THz peak electric field of $\mathcal{E}_A^0 + \mathcal{E}_B^0 = 11$ kV/cm as a function of $t$ and $\tau$. Lines of constant phase of pulse A appear as vertical lines, while those of pulse B appear along a 45° angle, since its field crests are located at constant values of $t$-$\tau$. At negative delay times when pulse A precedes pulse B, pulse A induces a polarization in the sample, which can oscillate freely until pulse B arrives. From there on, a nonlinear polarization emerges at $\nu_{ISB}$, owing to the interaction of the field of pulse B with the polarization triggered by pulse A. At $\tau = 0$, both pulses coincide temporally, giving rise to the strongest nonlinear response. For $\tau > 0$, pulses A and B change their roles, and the equivalent dynamics occur. While the real-time oscillation of the nonlinear polarization is directly traced along the $t$-axis, the coherent system memory manifests itself as a long-lived oscillatory signature along the $\tau$-axis. A slice through $\mathcal{E}_{NL}$ along the $\tau$-axis for a fixed delay time of $t = 0.5$ ps, thus, allows a first qualitative extraction of the polarization lifetime $T_2^*$, as well as the incoherent carrier lifetime $T_1$ (see Fig. 2(b)).

We model both components using the following fitting function:

$$\mathcal{E}_{NL}(\tau) = A_1 \cdot \exp\left(-\frac{\tau}{T_1}\right) + A_2 \cdot \sin(\omega\tau) \cdot \exp\left(-\frac{\tau}{T_2^*}\right), \quad (1)$$

which includes an exponential decay of the carrier population within the first term and an oscillatory exponential decay of the polarization within the second term. The fit is shown by the red dashed line in Fig. 2(b), with the decay of the carrier population (represented by the first term) shown by the solid black line. From this fit, we obtain a population lifetime of $T_1 = (5.0 \pm 0.3)$ ps and a coherence time of $T_2^* = (2.2 \pm 0.2)$ ps; this indicates a fast relaxation time in the saturable absorber, which is well within the gain recovery time of a THz QCL.

## 3. Liouville path analysis: role of coherent and incoherent nonlinearities

By performing a two-dimensional Fourier transform (Fig. 2(c)) of the time-domain data, we can rigorously decompose the total nonlinear response into all contributing nonlinear interaction processes [32]. The frequency axes $\nu_t$ and $\nu_\tau$ are associated with the electro-optic sampling time $t$, and the relative delay time $\tau$, respectively. Each spectrum is normalized to the spectral amplitude $\mathcal{E}_0$ of the driving field at the frequency of the ISB transition, $\nu_{ISB}$. In Fig. 2(d), four distinct maxima are observed at $(\nu_t, \nu_\tau) = (\nu_{ISB}, 0)$, $(\nu_{ISB}, -\nu_{ISB})$, $(\nu_{ISB}, \nu_{ISB})$, and $(\nu_{ISB}, -2\nu_{ISB})$. The position of each of these peaks can be uniquely expanded into a linear combination, using integer coefficients, of wavevectors of the incident fields (see arrows $k_A$, $k_B$ in Fig. 2(d)), representing a corresponding Liouville path. In particular, the peak at $(\nu_t, \nu_\tau) = (\nu_{ISB}, 0)$ can be expressed as $k_{PP1} = k_A + k_B - k_B$, characterizing an incoherent pump-probe signal (PP1), where the phase of pulse B cancels out. In this configuration, pulse B acts as a pump and pulse A as a probe. The maximum at $(\nu_t, \nu_\tau) = (\nu_{ISB}, -\nu_{ISB})$ is the equivalent pump-probe signal where pulses A and B switch roles (PP2). Both pump-probe interactions are the origin of the purely exponential decay in Fig. 2(b). In contrast, the four-wave mixing (4WM) signals at $(\nu_{ISB}, \nu_{ISB})$ and $(\nu_{ISB}, -2\nu_{ISB})$ contain wave-vector combinations that preserve the phases of both fields: the signal at $(\nu_{ISB}, -2\nu_{ISB})$ can be expressed as $k_{4WM2} = 2k_B - k_A$ and the signal at $(\nu_{ISB}, \nu_{ISB})$ as $k_{4WM1} = 2k_A - k_B$. Thus, these 4WM signals provide a direct way of monitoring the polarization of the electronic system.

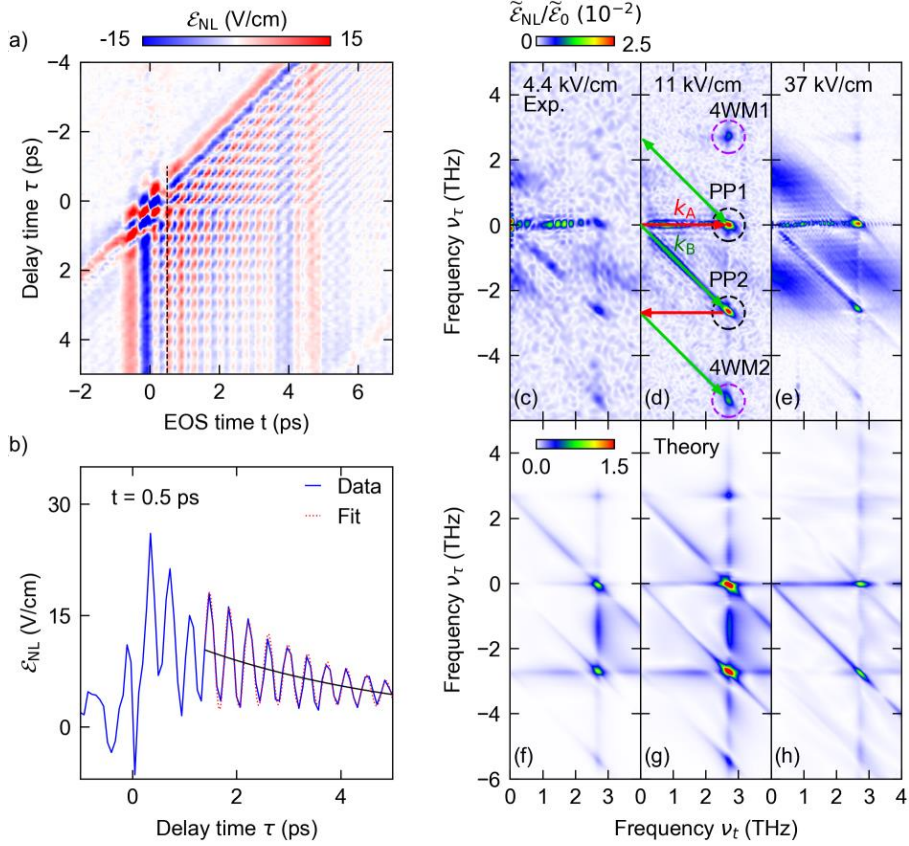

Fig. 2. (a) Nonlinear signal $\mathcal{E}_{NL} = \mathcal{E}_{AB} - \mathcal{E}_A - \mathcal{E}_B$ (see text) as a function of the electro-optic delay time $t$, and $\tau$, for peak incident electric fields of 11 kV/cm. (b) Slice of $\mathcal{E}_{NL}$ for a fixed time $t = 0.5$ ps (see black dashed line in (a)), revealing the decay of coherence (oscillatory component, $T_2^* = 2.2$ ps) and population (offset component, black curve, $T_1 = 5.0$ ps). (c)–(e) Fourier-transform $\tilde{\mathcal{E}}_{NL}$ of time-domain data $\mathcal{E}_{NL}$ for experimental field amplitudes of 4.4, 11, and 37 kV/cm, each normalized to the spectral amplitude $\mathcal{E}_0$ of the driving field at 2.7 THz. (d) Breakdown of nonlinear response into pump-probe (PP1, PP2: black circles) and four-wave mixing (4WM1, 4WM2: purple circles) contributions, and corresponding Liouville paths, each consisting of a superposition of the wave vectors of $\mathcal{E}_A$ ($k_A$, red arrows) and $\mathcal{E}_B$ ($k_B$, green arrows). (f)-(h) Theoretical nonlinear response calculated by numerically solving the optical Bloch equations.

We next investigate the amplitude scaling of these nonlinearities as a function of the peak amplitude of the driving field $\mathcal{E}_0$, varied from 4.4 kV/cm to 37 kV/cm in a series of 2D measurements. This amplitude range covers not only the onset of nonlinearities at low field amplitudes (Fig. 2(c)) but also the strong-field regime, which is accompanied by a reduction of the amplitude of both pump-probe and four-wave mixing signals. For a quantitative investigation of the field dependence of the different nonlinearities, the amplitudes of the PP and the 4WM signals were integrated within a frequency window of 0.5 THz around their respective center frequencies. Figure 3(a) shows the spectral weight of the PP2 and Fig. 3(b) the spectral weight of the 4WM1 signal as a function of the peak electric field (red crosses). Both curves show a sharp onset of the nonlinearity at low field strengths with a pronounced maximum at a field strength of 11 kV/cm, which is followed by a decrease in the integrated amplitude. While a purely incoherent saturation mechanism, e.g. by scattering and heating, would diminish the 4WM signal and leave the PP signal saturated, in our experiment both contributions decrease rapidly as $\mathcal{E}_0$ is increased further. We will show in the following that this behavior is a hallmark of an ultrafast THz-driven Rabi flopping [35] of the population on the time scale of the oscillation period of the THz wave and the intersubband transition.

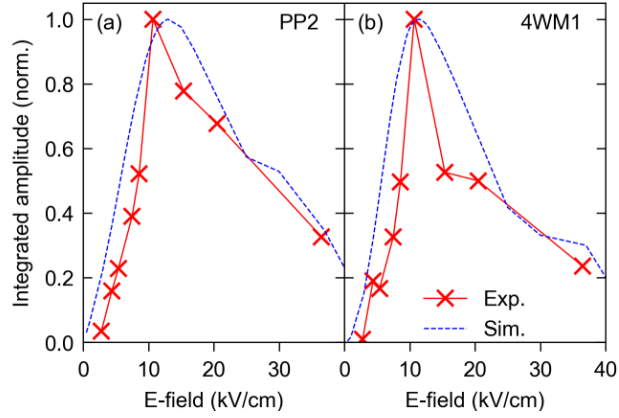

Fig. 3. Integrated spectral weight of (a) pump-probe (PP2) and (b) four-wave mixing (4WM1) signals as a function of the peak electric field of the driving THz pulses. Red crosses and lines: experiment; blue dashed curves: simulation.

## 4. Theoretical analysis of sub-cycle polarization dynamics

We employ a semiclassical theory based on the Maxwell-Bloch equations to simulate the polarization dynamics. Our theory goes beyond the rotating-wave-approximation and includes a self-consistent treatment of the re-emitted electric field [36]. The system's state is expressed by the density matrix $\rho$, whose temporal evolution is described by the von Neumann equation:

$$i\hbar \frac{\partial \rho}{\partial t} = [H, \rho] - i\gamma \circ \rho \qquad (2)$$

$$H = \begin{pmatrix} \hbar\omega_1 + \hbar\Omega_{11} & \hbar\Omega_{12} e^{-i\omega_{12}t} \\ \hbar\Omega_{12} e^{i\omega_{12}t} & \hbar\omega_2 + \hbar\Omega_{22} \end{pmatrix} \qquad (3)$$

where " $\circ$ " denotes the element-wise product. Here, the Hamiltonian H contains the intersubband energies $\hbar\omega_1$ and $\hbar\omega_2$; the resulting transition energy, $\hbar\omega_{12} = \hbar\omega_2 - \hbar\omega_1$; the Rabi frequency, $\Omega_{12} = \mu_{12} \mathcal{E}(t)/\hbar$; and, the ISB transition dipole matrix element $\mu_{12} = \langle \Psi_1 | e \cdot z | \Psi_2 \rangle$ (see Fig. 1(b)). The additional terms $\hbar\Omega_{ii} = \mu_{ii}\mathcal{E}(t)$ in the diagonal elements account for the permanent dipole moment $\mu_{ii}$ of the subband wave functions due to the asymmetry of the quantum wells. These dipoles are calculated from the simulated wave functions in Fig 1(b) as $\mu_{ii} = \langle \Psi_i | e \cdot z | \Psi_i \rangle$. The transmitted electric field $\mathcal{E}(t)$ is determined by a self-consistent treatment of both the incident electric field $\mathcal{E}_{\text{THz}}$ as measured separately, and the field emitted by the ISB polarization (second term):

$$\mathcal{E}(t) = \mathcal{E}_{\text{THz}} - \Gamma \mu_0 \frac{c}{2n} \frac{\partial \rho_{12}}{\partial t} \qquad (4)$$

Here $\mu_0$ denotes the vacuum permeability, $c$ is the speed of light in vacuum, and $n$ is the refractive index of the material. The constant $\Gamma = 1.7 \times 10^{13}$ describes the coupling of the near-fields of the electron ensemble to the far field by the grating, and is chosen to yield the best fit to the experiment.

Dephasing is introduced by a relaxation time approximation based on the experimentally determined decay constants:

$$\gamma = \begin{pmatrix} \frac{1}{T_1} & \frac{1}{T_2^*} \\ \frac{1}{T_2^*} & -\frac{1}{T_1} \end{pmatrix} \quad (5)$$

The spatial Gaussian shape of the THz electric field in the focal plane is taken into account by employing a weighted average over calculations for several peak field amplitudes.

Figures 2(f)-2(h) show the resulting 2D spectra calculated from this model, compared to the experimental data, shown in Fig. 2(c)-2(e). We find excellent agreement: Both the field dependence of the amplitude, as well as the spectral shape of the respective nonlinearities, are reproduced by the model. Our theory also allows us to explain the broadband tails of each nonlinear interaction as a consequence of the large bandwidth of our single-cycle THz pulses. While the high-frequency wing of the incident spectrum drives the ISB transition resonantly, the strong frequency components in the spectral center induce non-resonant transitions. The latter contribute broadband nonlinearities, even though they do not cause perfect population inversion. An additional, low-frequency response is contributed by the interaction of the THz electric field with the asymmetry-related permanent dipole moment.

For a quantitative comparison, we plot the integrated amplitude of the PP and 4WM signals as blue dashed curves in Figs. 3(a) and 3(b), which coincide well with the experimental data. For low fields, the nonlinearities rise steeply until they reach a maximum at a peak field of 11 kV/cm before they decay at yet higher field amplitudes. The different widths of the PP and 4WM maxima may be attributed to the fact that 4WM is inherently more susceptible to dephasing than incoherent PP signals. Furthermore, the 4WM signal saturates for π/2 pulses whereas the maximum PP signal occurs for π pulses. While these signatures are strongly averaged out by the spatial inhomogeneity of the near field, the effect contributes to the field scaling of the nonlinearities.

The good agreement between theory and experiment allows us to extract microscopic dynamics, such as the diagonal elements of the density matrix and their temporal evolution, which are not directly visible in the experiment. In Fig. 4(a), we investigate these dynamics and plot the population inversion $w = \rho_{22} - \rho_{11}$, as a function of the electro-optic delay time $t$ and the relative delay time $\tau$. Strong coherent dynamics occur on timescales shorter than $T_2^*$ leading to almost complete population inversion within one cycle of the optical carrier wave. For a delay time of $\tau = -0.094$ ps, pulse A prepares a population inversion, which is coherently increased to a value of $w = 0.9$ by pulse B within only 0.3 ps. This timescale is shorter than a single oscillation period of the transition frequency $\nu_{ISB}$.

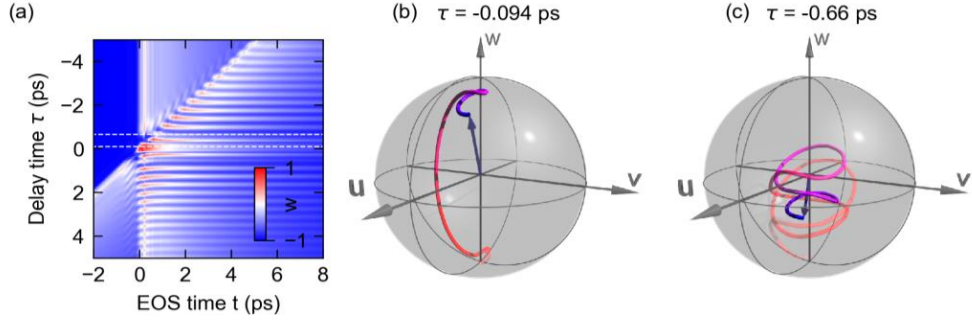

Fig. 4. (a) Calculated population inversion $w = \rho_{22} - \rho_{11}$ as a function of the delay times $t$ and $\tau$ for a peak electric field of 15 kV/cm. (b) Trajectory of the Bloch vector (**B**) for a delay time $\tau = -0.094$ ps and (c) $\tau = -0.66$ ps (indicated by white dashed lines in (a)). Red (blue) colors indicate early (late) EOS time $t$. u and v represent the real and imaginary part of the polarization, respectively.

This situation is illustrated in Fig. 4(b), where the trajectory of the complex Bloch vector $\mathbf{B} = (u, v, w)$ within the Bloch sphere is plotted and the polarization is mapped as $\rho_{12} = u + iv$. In contrast, for a delay time of $\tau = -0.66$ ps, pulse B coherently depopulates the upper level due to a phase shift of π relative to the excitation pulse, as shown in Fig. 4(c). The remaining population inversion after pulse B is due to dephasing of the system with the decay time $T_2^*$. Note that all dynamics occur in a regime where the Rabi frequency, the intersubband transition frequency and the carrier frequency of the THz field, all are of comparable size. This extreme limit of nonlinear light-matter interaction, termed carrier-wave Rabi flopping, has previously been studied only in the near-infrared spectral domain [35]. To the best of our knowledge, our results represent the first observation of this regime in low-frequency THz ISB transitions.

Performing the simulation for different THz waveforms gives an estimate of the peak fields necessary to invert the system, which is a crucial figure of merit for future applications. Assuming a resonant excitation with a 4 ps pulse centered on $\nu_{ISB}$, which is currently achieved using actively mode locked QCLs, a THz peak electric field of 1 kV/cm is already sufficient to drive the system into strong saturation. Although the far-field of QCLs is still in the range of tens of V/cm, the strongly confined waveguide mode inside the laser cavity reaches peak values of 150 V/cm. By finely tuning our system parameters, such as the doping concentration, the thickness of the quantum wells, or by optimizing the coupling of the QCL mode to the electronic wave function through improved metallic nanostructures, these field amplitudes should already saturate the ISB transition, making the proposed MQW structures a promising candidate for use as a saturable absorber in QCLs.

## 5. Conclusions

In conclusion, we have investigated the nonlinear dynamics of the intersubband transition in a MQW system under strong resonant THz excitation. Coherent and incoherent contributions to the nonlinear signal have been identified as pump-probe and four-wave-mixing processes, enabling us to trace the population and nonlinear polarization of intersubband electrons with sub-cycle resolution. Evaluating their dynamics revealed relaxation times of the MQW system of 5 ps, significantly shorter than typical gain relaxation times of QCLs. These nonlinearities showed a clear saturation behaviour with increasing THz peak electric field, representing the first 2D spectroscopic observation of THz carrier-wave Rabi flopping. We quantitatively reproduced our results by a numerical model based on a self-consistent solution of the Maxwell-Bloch equations for a two-level system beyond the rotating-wave approximation. In particular, the calculation reveals a population of the upper subband generated by the first pulse, followed by a coherent depopulation by the second THz pulse. Our experiment and theory provide a deep understanding of the sub-cycle dynamics of intersubband transitions and highlight a route for designing customized heterostructures implementing saturable absorbers in the THz range. Corresponding future QCL structures may allow passive mode locking and thus open up the perspective of generating high-power, widely tunable ultrashort THz pulses from electrically biased semiconductor devices.

**Funding**

Horizon 2020 Framework Programme (665158); Engineering and Physical Sciences Research Council (EP/P021859/1, EP/M028143/1).

**Acknowledgments**

We thank Imke Gronwald for technical support. EHL acknowledges support from the Royal Society and the Wolfson Foundation. M.S.V. acknowledges partial support from the second half of the Balzan Prize 2016 in applied photonics delivered to Federico Capasso.

**References**

1. R. Ulbricht, E. Hendry, J. Shan, T. F. Heinz, and M. Bonn, "Carrier dynamics in semiconductors studied with time-resolved terahertz spectroscopy," Rev. Mod. Phys. **83**, 543–586 (2011).
2. P. U. Jepsen, D. G. Cooke, and M. Koch, "Terahertz spectroscopy and imaging - Modern techniques and applications," Laser & Photon. Rev. **5**, 124–166 (2011).
3. Y. C. Shen, P. C. Upadhya, H. E. Beere, E. H. Linfield, A. G. Davies, I. S. Gregory, C. Baker, W. R. Tribe, and M. J. Evans, "Generation and detection of ultrabroadband terahertz radiation using photoconductive emitters and receivers," Appl. Phys. Lett. **85**, 164–166 (2004).
4. A. G. Stepanov, J. Hebling, and J. Kuhl, "Efficient generation of subpicosecond terahertz radiation by phase-matched optical rectification using ultrashort laser pulses with tilted pulse fronts," Appl. Phys. Lett. **83**, 3000–3002 (2003).
5. M. D. Thomson, M. Kreß, T. Löffler, and H. G. Roskos, "Broadband THz emission from gas plasmas induced by femtosecond optical pulses: from fundamentals to applications," Laser & Photon. Rev. **1**, 349–368 (2007).
6. O. Schubert, M. Hohenleutner, F. Langer, B. Urbanek, C. Lange, U. Huttner, D. Golde, T. Meier, M. Kira, S. W. Koch and R. Huber, "Sub-cycle control of terahertz high-harmonic generation by dynamical Bloch oscillations," Nature Photonics **8**, 119-123 (2014).
7. T. Kampfrath, A. Sell, G. Klatt, A. Pashkin, S. Mährlein, T. Dekorsy, M. Wolf, M. Fiebig, A. Leitenstorfer, and R. Huber, "Coherent terahertz control of antiferromagnetic spin waves," Nat. Photon. **5**, 31–34 (2011).
8. F. Junginger, B. Mayer, C. Schmidt, O. Schubert, S. Mährlein, A. Leitenstorfer, R. Huber, and A. Pashkin, "Non-perturbative interband response of InSb driven off-resonantly by few-cycle electromagnetic transients," Phys. Rev. Lett. **109**, 147403 (2012).
9. J. Lu, X. Li, H. Y. Hwang, B. K. Ofori-Okai, T. Kurihara, T. Suemoto, and K. Nelson, "Coherent two-dimensional terahertz magnetic resonance spectroscopy of collective spin waves," Phys. Rev. Lett. **118**, 207204 (2017).
10. S. S. Dhillon, M. S. Vitiello, E. H. Linfield, A. G. Davies, M. C. Hoffmann, J. Booske, C. Paoloni, M. Gensch, P. Weightman, G. P. Williams, E. Castro-Camus, D. R. S. Cumming, F. Simoens, I. Escorcia-Carranza, J. Grant, S. Lucyszyn, M. Kuwata-Gonokami, K. Konishi, M. Koch, C. A. Schmuttenmaer, T. L. Cocker, R. Huber, A. G. Markelz, Z. D. Taylor, V. P. Wallace, J. A. Zeitler, J. Sibik, T. M. Korter, B. Ellison, S. Rea, P. Goldsmith, K. B. Cooper, R. Appleby, D. Pardo, P. G. Huggard, V. Krozer, H. Shams, M. Fice, C. Renaud, A. Seeds, A. Stöhr, M. Naftaly, N. Ridler, R. Clarke, J. E. Cunningham, and M. B. Johnston, "The 2017 terahertz science and technology roadmap," J. Phys. D: Appl. Phys. **50,** 43001–43049 (2017).
11. R. Köhler, A. Tredicucci, F. Beltram, H. E. Beere, E. Linfield, A. G. Davies, D. A. Ritchie, R. C. Iotti, and Fausto Rossi, "Terahertz semiconductor-heterostructure laser," Nature **417**, 156–159 (2002).
12. J. Kröll, J. Darmo, S. S. Dhillon, X. Marcadet, M. Calligaro, C. Sirtori, and K. Unterrainer, "Phase-resolved measurements of stimulated emission in a laser," Nature **449**, 698–701 (2007).
13. D. Oustinov, N. Jukam, R. Rungsawang, J. Madéo, S. Barbieri, P. Filloux, C. Sirtori, X. Marcadet, J. Tignon, and S. Dhillon, "Phase seeding of a terahertz quantum cascade laser," Nat. Commun. **1**, 69 (2010).
14. M. S. Vitiello, G. Scalari, B. S. Williams, P. De Natale, "Quantum cascade lasers: 20 years of challenges," Optics Express **23**, 5167–5182 (2015).
15. M. Rösch, G. Scalari, M. Beck, and J. Faist, "Octave-spanning semiconductor laser," Nat. Photon. **9**, 42–47 (2015).
16. H. Li, P. Laffaille, D. Gacemi, M. Apfel, C. Sirtori, J. Leonardon, G. Santarelli, M. Rösch, G. Scalari, M. Beck, J. Faist, W. Hänsel, R. Holzwarth, and S. Barbieri, "Dynamics of ultra-broadband terahertz quantum cascade lasers for comb operation," Opt. Express **23**, 33270–33294 (2015).
17. J. Faist, G. Villares, G. Scalari, M. Rösch, C. Bonzon, A. Hugi, and M. Beck, "Quantum cascade laser frequency combs," Nanophotonics, **5**(2), 272-291 (2016).

18. S. Barbieri, M. Ravaro, P. Gellie, G. Santarelli, C. Manquest, C. Sirtori, S. P. Khanna, E. H. Linfield, and A. G. Davies, "Coherent sampling of active mode-locked terahertz quantum cascade lasers and frequency synthesis," Nat. Photon. **5**, 306–313 (2011).
19. F. Wang, K. Maussang, S. Moumdji, R. Colombelli, J. R. Freeman, I. Kundu, L. Li, E. H. Linfield, A. G. Davies, J. Mangeney, J. Tignon, and S. S. Dhillon, "Generating ultrafast pulses of light from quantum cascade lasers," Optica **2**, 944-949 (2015).
20. F. Wang, H. Nong, T. Fobbe, V. Pistore, S. Houver, S, Markmann, N. Jukam, M. Amanti, C. Sirtori, S. Moumdji, R. Colombelli, L. Li, E. Linfield, G. Davies, J. Mangeney, J. Tignon, and S. Dhillon, "Short terahertz pulse generation from a dispersion compensated modelocked semiconductor laser ," Laser & Photon. Rev. **11**, 1700013 (2017).
21. A. Mottaghizadeh, D. Gacemi, P. Laffaille, H. Li, M. Amanti, C. Sirtori, G. Santarelli, W. Hänsel, R. Holzwart, L. H. Li, E. H. Linfield, and S. Barbieri, "5-ps-long terahertz pulses from an active-mode-locked quantum cascade laser," Optica **4**, 168-171 (2017).
22. U. Keller, D. A. B. Miller, G. D. Boyd, T. H. Chiu, J. F. Ferguson, and M. T. Asom, "Solid-state low-loss intracavity saturable absorber for Nd:YLF lasers: an antiresonant semiconductor Fabry–Perot saturable absorber," Opt. Lett. **17**, 505-507 (1992).
23. M. C. Hoffmann, and D. Turchinovich, "Semiconductor saturable absorbers for ultrafast terahertz signals," Appl. Phys. Lett. **96**, 151110 (2010).
24. V. Bianchi, T. Carey, L. Viti, L. Li, E. H. Linfield, A. G. Davies, A. Tredicucci, D. Yoon, P. G. Karagiannidis, L. Lombardi, F. Tomarchio, A. C. Ferrari, F. Torrisi, and M. S. Vitiello, "Terahertz saturable absorbers from liquid phase exfoliation of graphite," Nat. Commun. **8**, 15763 (2017).
25. J. N. Heyman, K. Unterrainer, K. Craig, J. Williams, M. S. Sherwin, K. Campman, P. F. Hopkins, A. C. Gossard, B. N. Murdin, and C. J. G. M. Langerak, "Far-infrared pump-probe measurements of the intersubband lifetime in an AlGaAs/GaAs coupled-quantum well," Appl. Phys. Lett. **68**, 3019–3021 (1996).
26. F. Eickemeyer, M. Woerner, A. M. Weiner, and T. Elsaesser, "Coherent nonlinear propagation of ultrafast electric field transients through intersubband resonances," Appl. Phys. Lett. **79**, 165–167 (2001).
27. T. Müller, W. Parz, G. Strasser, and K. Unterrainer, "Pulse-induced quantum interference of intersubband transitions in coupled quantumwells," Appl. Phys. Lett. **84**, 64–66 (2004).
28. G. Folpini, D. Morrill, C. Somma, K. Reimann, M. Woerner, T. Elsaesser, and K. Biermann, "Nonresonant coherent control: Intersubband excitations manipulated by a nonresonant terahertz pulse," Phys. Rev. B **92**, 85306–85313 (2015).
29. D. Dietze, J. Darmo, and K. Unterrainer, "THz-driven nonlinear intersubband dynamics in quantum wells," Opt. Express **20**, 23053-23060 (2012).
30. C. W. Luo, K. Reimann, M. Woerner, T. Elsaesser, R. Hey, and K. H. Ploog, "Phase-resolved nonlinear response of a two-dimensional electron gas under femtosecond intersubband excitation," Phys. Rev. Lett. **92**, 047402 (2004).
31. S. Birner, T. Zibold, T. Andlauer, T. Kubis, M. Sabathil, A. Trellakis, and P. Vogl, "Nextnano: general purpose 3-D simulations," IEEE Trans. Electron Devices, **54**, 2137–2142, (2007).
32. W. Kuehn, K. Reimann, M. Woerner, and T. Elsaesser, "Phase-resolved two-dimensional spectroscopy based on collinear n-wave mixing in the ultrafast time domain," J. Chem. Phys. **130**, 164503 (2009).
33. T. Maag, A. Bayer, S. Baierl, M. Hohenleutner, T. Korn, C. Schüller, D. Schuh, D. Bougeard, C. Lange, R. Huber, M. Mootz, J. E. Sipe, S. W. Koch, and M. Kira, "Coherent cyclotron motion beyond Kohn's theorem," Nat. Phys. **12**, 119–123 (2016).
34. S. Markmann, H. Nong, S. Pal, T. Fobbe, N. Hekmat, R. A. Mohandas, P. Dean, L. Li, E. H. Linfield, A. G. Davies, A. D. Wieck, and N. Jukam, "Two-dimensional coherent spectroscopy of a THz quantum cascade laser: observation of multiple harmonics," Opt. Express **25**, 21753-21761 (2017).
35. O. D. Mücke, T. Tritschler, and M. Wegener, "Signatures of carrier-wave rabi flopping in GaAs," Phys. Rev. Lett. **87**, 057401 (2001).
36. M. Kira, S. W. Koch, "Many-body correlations and excitonic effects in semiconductor spectroscopy," Prog. Quantum Electron. **30**, 155–296 (2006).